      [1999/12/01 v1.4c Il Nuovo Cimento]
\documentclass{cimento}

\usepackage{graphicx}
\begin{document}
\title{Analysis of the BATSE Continuous MER Data}
\author{
P.~Veres\from{ins:x}, I.~Horv\'ath\from{ins:y} \atque
L.G.~Bal\'azs\from{ins:z}}
\instlist{\inst{ins:x} E\"otv\"os University, Budapest, P\'azm\'any P\'eter s\'et\'any 1/A, H-1518, Budapest, Hungary

\inst{ins:y} Department of Physics, Bolyai Military Univ. POB 12,
H-1456 Budapest,  Hungary

  \inst{ins:z} Konkoly Observatory, POB 67, H-1525 Budapest, Hungary }

\PACSes{\PACSit{98.70.Rz (GRB, 
gamma ray sources)}
{98.80.-k (cosmology)}}

\maketitle
\begin{abstract}
The CGRO/BATSE database includes many types of data such as the 16-channel
continuous background or medium energy resolution burst data (CONT and MER data
types). We have calculated some four hundred burst's medium energy resolution
spectra and Principal Component Analysis has been applied. We found five
components can describe GRBs' spectra.
\end{abstract}

\section{Introduction}
In the process of understanding the internal mechanism of the
gamma ray bursts it is of great importance, to classify the
events, to find important and less important variables among the
observed parameters. Searching for groups and subgroups is still
one of the main goals of the GRB research. The BATSE GRB database
produced interesting results (e.g.
\cite{ref:hak,ref:ho1,ref:ho2,ref:mbv,ref:muk} and the references
therein). BATSE CONT data are available on the Internet via
anonymous ftp from the CGRO Science Support Center site
\cite{ref:batse}. The CONT data type is trigger independent, has a
spectral resolution of $16$ channel and a temporal resolution of
$2.048$ s. Data are organized into 8x16 matrices for each time
bin, 8 being the number of detectors and 16 the number of energy
channels. These channels are used to approximate the burst's
spectrum which is quite variable during the burst (e.g.
\cite{ref:me1,ref:ry1}). We estimated using Principal Component
Analysis (PCA) how many parameters are necessary to characterize
CONT data.

\section{Data Selection}
In the CONT database there are $\sim 900$ bursts ranging from
BATSE trigger number $105$ to $3192$. The data are
background-subtracted and continuous. There is no clear indication
where the burst starts or where it ends, so we defined the
following criteria: If the count rate exceeds the $4 \sigma$ level
for a given energy channel, the entire bin is included in the
summation, where $\sigma$ is the variation of counts in the given
channel. The background variation of a channel ($\sigma$) is the
average count rate over selected "burst-free" intervals calculated
from 200 time bins when no burst has occurred. After the summation
every burst was represented by a $16$-element vector. At some
bursts none of the channels had a greater count rate than $4
\sigma $ and obviously these bursts were excluded from the sample.

\begin{table}
\caption{The eigenvalues of the 16 channel spectra. }
  \label{table:pca}
$$
\begin{array}{lll}
\hline \mbox{Component} & \mbox{Eigenvalue} & \mbox{Cumulative} \\
&  &  \mbox{percentage}   \\

\hline \textbf{1} & \textbf{6.12} & \textbf{38.2} \\
\textbf{2} & \textbf{3.32} & \textbf{59.0}   \\
 \textbf{3} & \textbf{1.34} & \textbf{67.4}   \\
 \textbf{4} & \textbf{0.89} & \textbf{73.0}   \\
 \textbf{5} & \textbf{0.82} & \textbf{78.1}   \\
 6 & 0.58 &   \\
 7 & 0.54 &   \\
 8 & 0.43 &   \\
 9 & 0.40 &   \\
 10 & 0.32 &  \\
 11 & 0.29 &  \\
 12 & 0.27 &  \\
 13 & 0.24 &  \\
 14 & 0.23 &  \\
 15 & 0.15 &  \\
 16 & 0 &     \\
\hline
\end{array}
$$
\end{table}

\section{Data processing}

Summing up the selected time bins (for the triggered detectors
only) we got a 16-element vector for each burst. Excluding the
bursts with no counts above the $4 \sigma$ level meant to exclude
$200$ GRBs. Further $250$ were excluded because they had more than
four negative elements. Therefore $401$ bursts remained in this
analysis. To proceed further in processing the data we took  the
logarithm of the count values and divided them with the sum of the
16 elements. (If we keep only the bursts with 16 positive values
only 104 bursts remain, therefore if the burst has less than five
negative values these values are replaced with 1, which should not
cause significant changes in our analysis.)

\begin{figure}[!ht]
\begin{center}
\scalebox{0.8}[0.8]{\includegraphics[width=\textwidth]{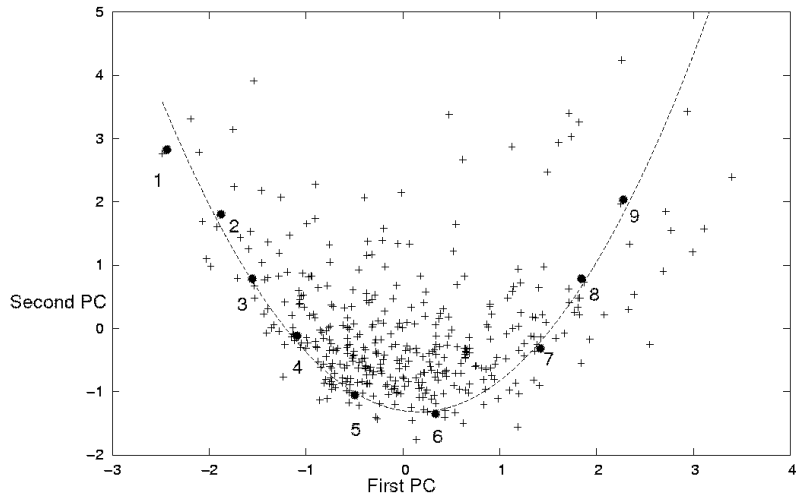}}

\caption{The first vs. second PCs distribution of the 401 bursts
and the one parameter fit. Dots represent the different spectral classes (see Fig. 2.). }

\end{center}
\end{figure}

\section{Principal Component Analysis}

The PCA \cite{ref:jolliffe} method has a  growing importance in
astronomy \cite{ref:pca,ref:balazs}. It is used to distinguish
between relevant and non-relevant variables. One can imagine the
method as a rotation in a multi-dimensional space (with normalized
variables) which finds the direction of the greatest standard
deviation. Based on our sample of $401$ bursts we have performed a
Principal Component Analysis. Since we divided with the sum of the
16 elements of the vector, the most obvious Principal Component,
the intensity of the burst has not been taken into account. Table
\ref{table:pca}. shows the eigenvalues of the PCA. If the
eigenvalue is less than 0.7 (Joliffe level \cite{ref:jolliffe})
the PC is not important. If the eigenvalue is bigger than 1 the PC
is very important. Therefore the five biggest PCs are enough to
describe the MER spectra of the bursts. These five PCs contain the
78.1\% of the whole MER spectral  information. {\bf Figure 1.}
shows the first vs. second PCs distribution. The 401 positions of
the bursts lay in an area that can be delimited by an arc. As {\bf
Figure 2.} shows the spectra change continuously along the arc.

\begin{figure}[!ht]
\begin{center}
\includegraphics[width=0.9\textwidth]{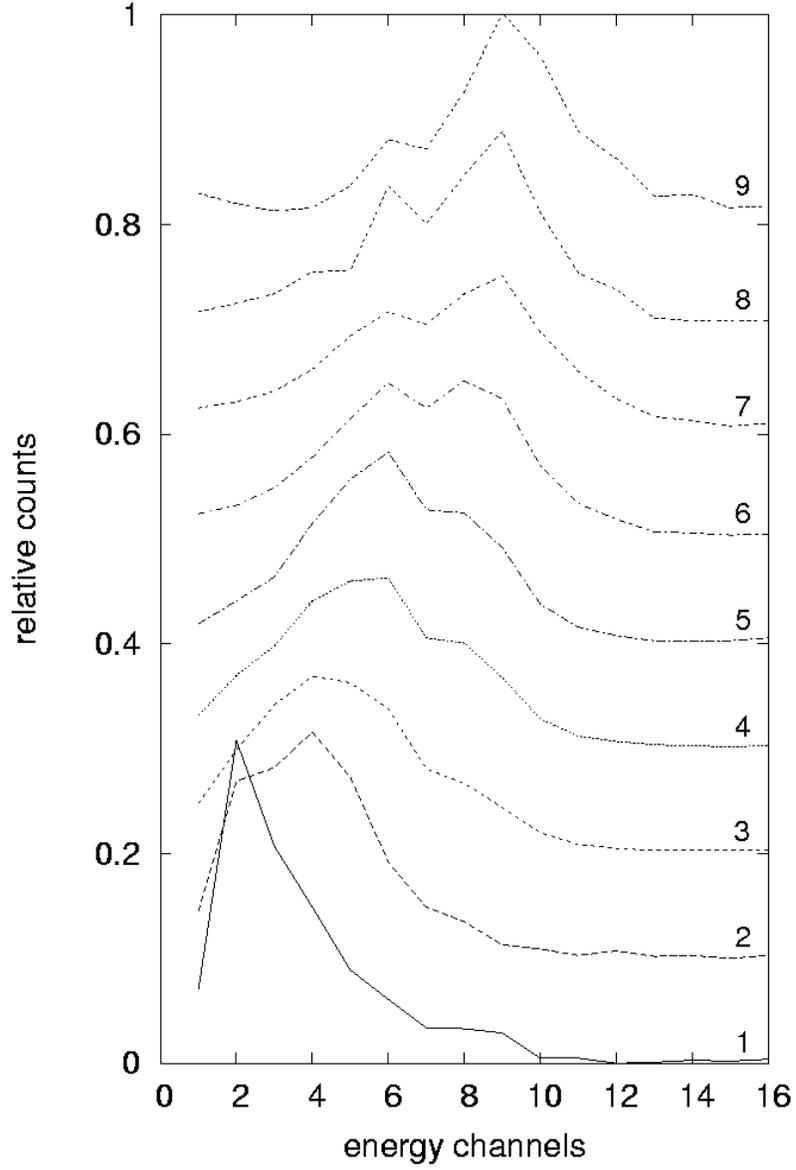}

\vspace{1mm}
\caption{Sample spectra. It can be easily seen that along the arc (see Fig. 1.)
the bursts' spectra change continuously from soft to hard. It is also
interesting to observe that there is a second peak which is barely visible in
the first position and it grows higher than the first peak as we gradually go
from position $1$ to position $9$.
}\label{figure:iv}
\end{center}
\end{figure}

\section{Conclusion}
We performed PCA on the 16x401 element matrix and we concluded
that the information specific for the spectra (not considering the
intensity) can basically be given by the first five Principal
Components. Furthermore, we recognized the distribution of GRBs in
the scatter plot of the first two PCs  can be delimited roughly by
an arc.

\acknowledgments

 This research was supported through OTKA grants T034549
and T48870.

\end{document}